# Resonant Inverse Photoelectron Spectroscopy of Ce and U Oxides

JG Tobin
U. Wisconsin-Oshkosh, Oshkosh, WI 54937
Email: tobinj@uwosh.edu

**Abstract**

The soft x-ray Resonant Inverse Photoelectron Spectroscopy (RIPES) of the $M_5$ transition in cerium oxide and the $N_5$ transition in uranium dioxide and supporting instrumentation is discussed. Further experimental evidence and analysis is presented to support the hypothesis that the Ce 5d states are intrinsic to the resonance process in cerium oxide. The strong connection with cerium dioxide is emphasized.

## I Introduction

Following the lead of Dose and coworkers [1,2], many new studies utilizing Inverse Photoelectron Spectroscopy at low energies (~ 10 eV) were pursued. [3 – 18] These studies included investigations of rare earth (4f) and actinide (5f) materials. [11, 14-18] In parallel with this, experimentation at higher energies (~1500 eV), i.e. Bremsstrahlung Isochromat Spectroscopy or BIS, was also carried out upon 4f and 5f materials. [18-23] The subject of this work is a series of Resonant Inverse Photoelectron Spectroscopy (RIPES) experiments performed at intermediate energies (hv ~ 800 eV) at the Ce $M_5$ and U $N_5$ edges of oxidized forms of these two elements and the supporting BIS, X-ray Emission Spectroscopy (XES) and instrument development. [24-40]

Interestingly, what is potentially the first RIPES measurements upon a 4f material was carried out by Chamberlain, Burr and Liefeld in 1974 [41, 42], utilizing a vacuum, two-crystal monochromator and a flowing gas proportional counter for the photon detection and an x-ray tube as the electron source. It seems likely that the samples were oxidized. In any case, it is clear that there is a beautiful resonance at the $M_{4,5}$ edges of the La and Ce samples. A small subset of their results is reproduced in Figure 1. Subsequently, there was substantial work on Resonant Inverse Photoemission in the Rare Earths, such as the investigations of Ce by Grioni et al. [43 -46]

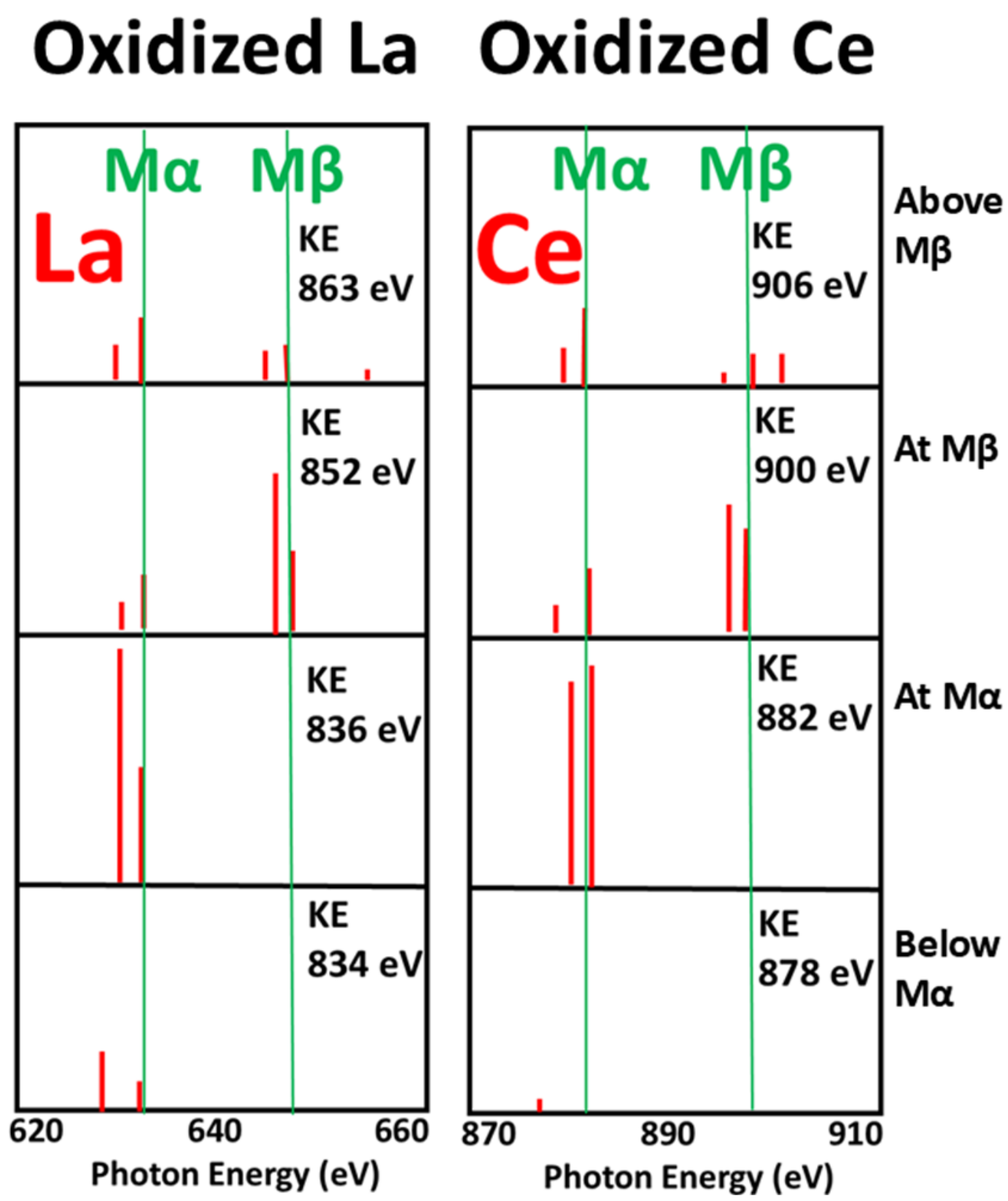

*Figure 1*
*Shown here is a subset of the data for the original RIPES measurements by Chamberlain, Burr and Liefeld. The red histograms are used to represent the peak positions and heights in the original spectra. Note the strong resonant behavior at the Mα and Mβ energies. Data extracted from Ref [41, 42].*

The results of the uranium dioxide RIPES experiment are summarized in Figure 2. It should be noted that our uranium dioxide RIPES experiment was part of a larger effort utilizing several different approaches, including X-ray Absorption Spectroscopy (XAS), BIS, XES and theoretical simulations. [25, 26, 30-32, 34, 35] In Panel B of Figure 2, the central experimental effect is presented: There are two RIPES maxima that directly correspond to two major peaks in the $N_5$ XES spectrum. These XES peaks persist as the dominant features over an extended excitation or Kinetic Energy (KE) range, from KE = 1000 eV to almost 4000 eV. [26, 29] It should also be noted that neither of these peaks is the $N_4$ feature: It is at hν ~ 775 eV. While the extraordinarily strong spectral feature is in XES, the explanation comes from RIPES. One of the central tenets of RIPES, as in Resonant Photoemission, is that both the direct and indirect channels must have the same final state. This inexorably leads to the conclusion that the key step in the indirect process of the satellite peak must involve a two-electron event. This hypothesis was confirmed by the observation that the main and satellite peaks are separated ~14 eV, which is two times the separation of the U6d and U5f manifolds. [25, 26, 30, 34]

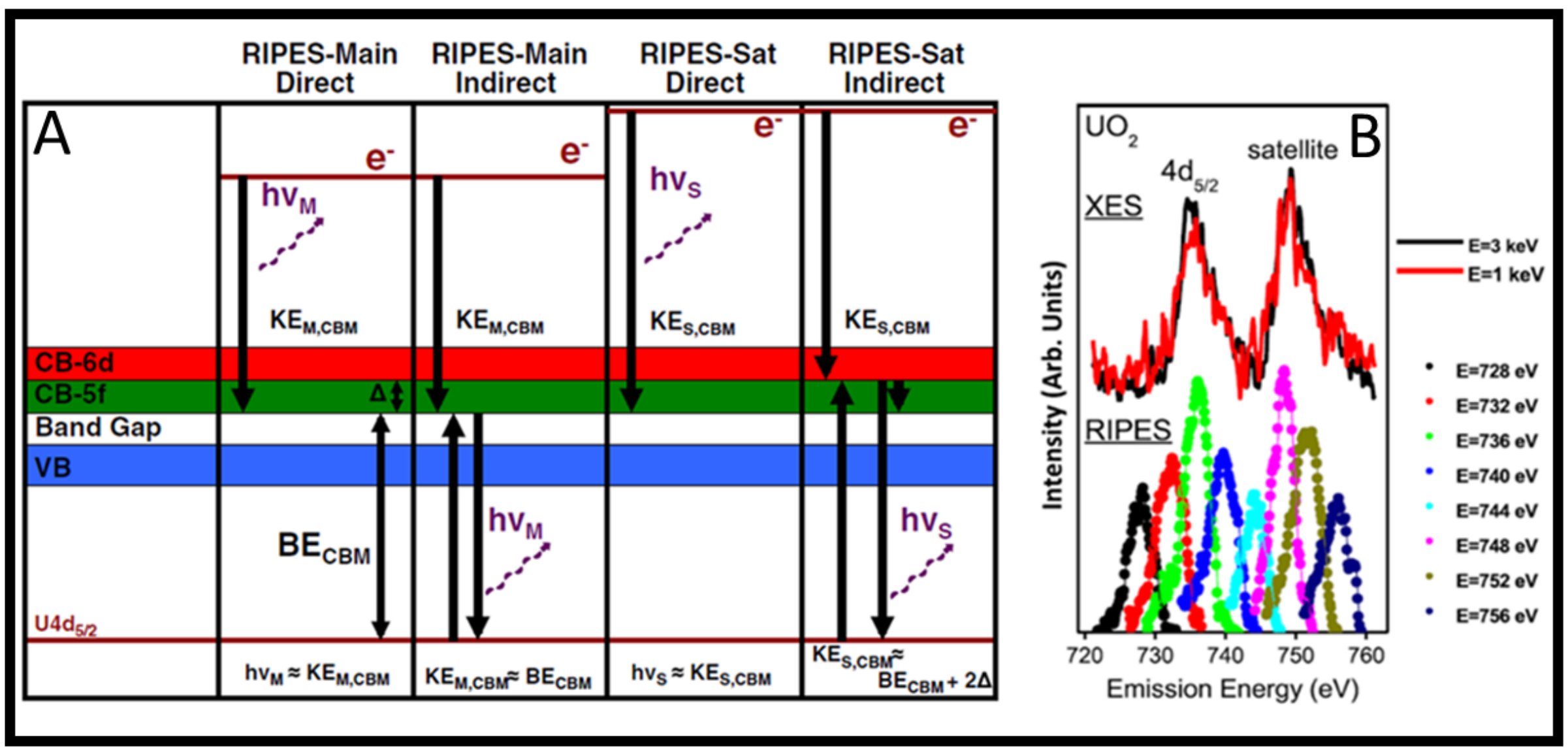

*Figure 2  Shown here is a synopsis of the results of the RIPES experiment upon uranium dioxide. A: An energy level diagram illustrating the direct and indirect processes. B: The U $N_5$ RIPES and XES spectra. Note the two maxima in the RIPES and the corresponding doublet in the XES. See the text for further discussion. VB is valence band. CB is conduction band. CBM is conduction band minimum. KE is kinetic energy.*

Before going on to a discussion of the cerium oxide results, two more points must be made, concerning uranium dioxide versus uranium metal. The first is that uranium metal has a different and much weaker resonant behavior. [31, 43] The second is that the separation of the U5f and U6d manifolds occurs in the dioxide [25, 26, 30, 34], but not the metal. [47]

# Resonant Inverse Photoelectron Spectroscopy of Ce and U Oxides

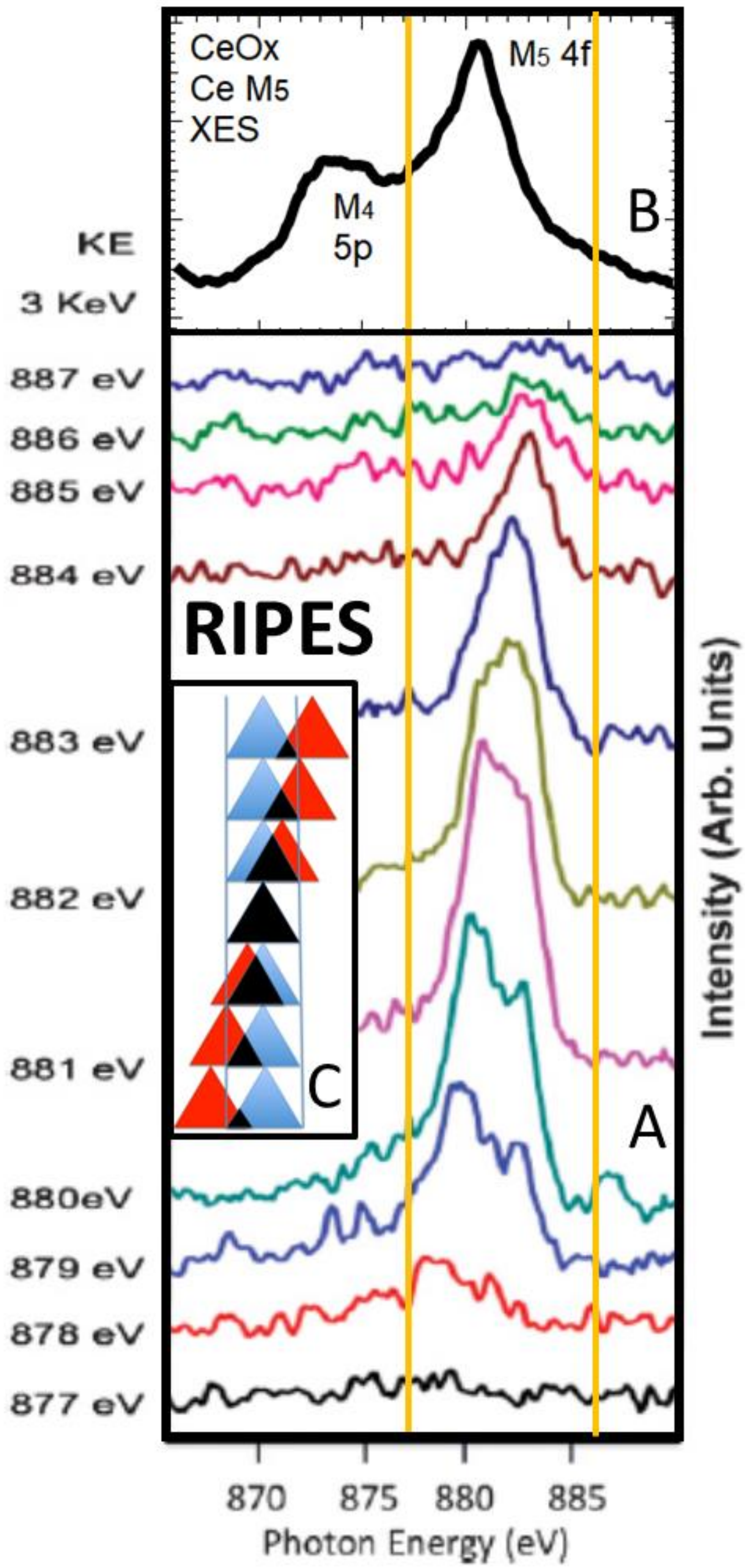

*Figure 3*
*A partial recap of the RIPES results for cerium oxide is presented here. A: the experimental RIPES spectra for the $M_5$ edge. KE is the electron gun kinetic energy. B: the XES spectra for the cerium oxide only, in the region near the Ce $M_5$ 4f peak and containing also the Ce $M_4$ 5p peak. C; a diagram illustrating the overlap model, where all of the RIPES intensity (black) comes from the overlap of the direct (red) and indirect (blue) channels. See text for further discussion.*

The cerium oxide results are shown in Figure 3. Here, the spectral results for the cerium oxide $M_5$ RIPES (Panel A) and XES (Panel B) are presented, as well as a schematic of a model for the RIPES involving the overlap (Panel C) in energy space of the direct and indirect channels (Figure 2). Again, this summary figure is built upon a series of experiments and analyses. [24, 27, 28, 33] In this work, the focus will be upon the results at the $M_5$ edge, for several reasons: it is most efficacious to compare the Ce $M_5$ to the U $N_5$; there is a strong match between the Ce $M_5$ RIPES and the cerium-oxide-only Ce $M_5$ XES; and the Ce $M_5$ edge exhibits the strongest RIPES effects. There is a RIPES effect exhibited at the Ce $M_4$ edge, but the effects are smaller and the alignment with the XES less clear.

The salient characteristic of the cerium oxide Ce $M_5$ RIPES is that there is only intensity in the energy regime near the $M_5$ edge. There is very little intensity below the $M_5$ or above the $M_4$ energies. This is very similar to what was observed earlier by Chamberlain, Burr and Liefeld. [41, 42] As can be seen in Figure 3, all of the intensity is in a box defined by the two vertical orange lines. The intensity variations can be easily explained by a simple model where the spectral intensity comes from an overlap of the direct and indirect channels (Figure 3, Panel C) Additionally, there are not two resonant features but only one. The combination of these observations leads one to consider the possibility that the single Ce $M_5$ peak is driven by something like the Indirect Channel in the Satellite Column in Figure 2, but for the Ce5d instead of the U6d.

The question then becomes: How would one test this? Because the there is no intensity other than the resonant peak, it is not feasible to measure the peak separations. However, in the sections below, two other possible spectroscopic pieces of evidence will be considered, that support the hypothesis of the importance of the Ce 5d states in the RIPES process.

The remainder of this work will be divided into three sections: II Experimental and Computational, III Results and Discussion and IV Conclusions and Summary. Because of the method of sample preparation, growing an oxide film on a Ce metal substrate, the exact nature of the oxide layer was unclear. Initially, it was thought that the oxide layer was $Ce_2O_3$. [24] Subsequent analysis of the XES results suggest instead that it is $CeO_2$. [33] For the sake of ease and simplicity, the spectroscopic modelling of the cerium oxide will use a $CeO_8$ cluster consistent with $CeO_2$ and the comparisons to more sophisticated computational modelling will emphasize $CeO_2$ as well.

## II Experimental and Computational

The RIPES experiments were performed at Lawrence Livermore National Laboratory (LLNL) using a spectrometer that went through a somewhat convoluted development. [36-40] As shown in Figure 4, the key components were a SPECS electron gun [26] and a soft-x-ray grating monochromator. [40] The monochromator was a commercially available XES 350 system from VG Scienta Ltd. Spectra were collected in “normal mode,” where the electron gun KE and the energy position of the center of the channel plate were both fixed and the energy distribution in the photon (hv) spectrum was derived from the intensities distributed across the channel-plate detector in the energy dispersal direction. (Figure 4, Panel C) Energy scale calibration was performed by direct measurement of the location of the energy threshold over a wide series of photon energies (Figure 4, Panel D) and a type of linear regression analysis upon the data. [26].

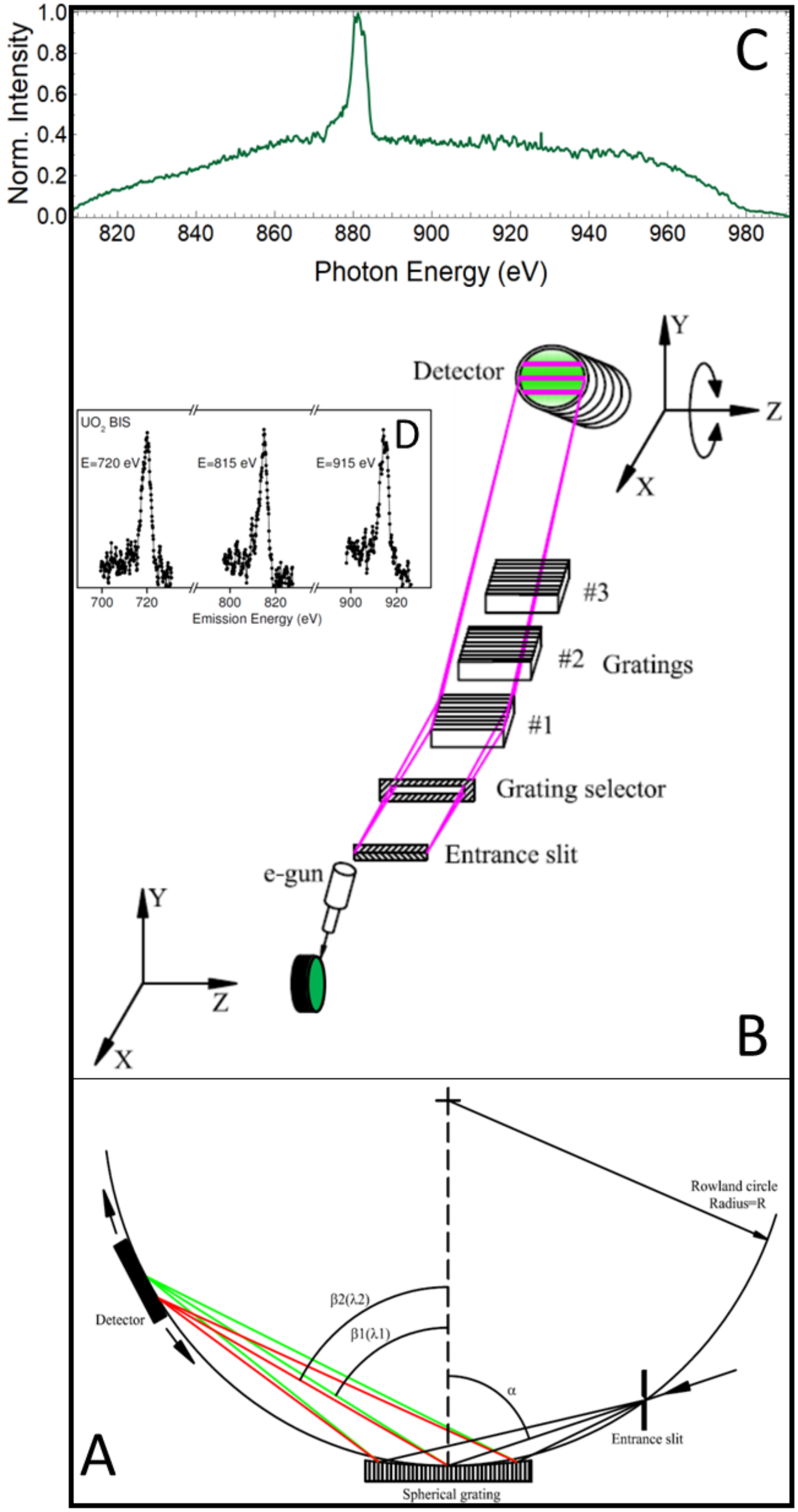
C
Norm. Intensity
1.0
0.8
0.6
0.4
0.2
0.0
820
840
860
880
900
920
940
960
980
Photon Energy (eV)
Detector
Y
Z
X
D
$UO_2$ BIS
E=720 eV
E=815 eV
E=915 eV
700
720
800
820
900
920
Emission Energy (eV)
#3
#2
Gratings
#1
Grating selector
Entrance slit
e-gun
Y
Z
X
B
Rowland circle
Radius=R
β2(λ2)
β1(λ1)
Detector
α
Entrance slit
A
Spherical grating

Figure 4

*Figure 4 (above) The instrumental details of the RIPES experiment are shown here. A: the Rowland Circle optics of the monochromator, including the entrance slit, grating and multichannel detector. [40] B: the experimental set-up with the sample, electron gun and monochromator. [40] C: an example of the output of the multichannel detector, for a single excitation kinetic energy. D: several BIS spectra of $UO_2$ used to calibrate the monochromator energy scale. [29] In Panel C, note the roll-off in intensity at the edges of the channel plate.*

As part of the analysis in the subsequent sections, the spectral analysis program FEFF will be used. FEFF is a data analysis platform used in x-ray absorption spectroscopy and associated techniques. It includes self-consistent real space multiple-scattering with simultaneous calculations of x-ray absorption spectra and the electronic structure, with special emphasis on the extended x-ray absorption fine structure (EXAFS). [48-51] The name FEFF originated with $f_{eff}$ or f-effective, a term from x-ray scattering theory. FEFF's beginnings trace directly to the requirement to quantify the elastic electron scattering events in the EXAFS final state, giving it a sensitivity to interatomic distances. The JFEFF version of FEFF 9.9.1 was used for these calculations. FEFF is a spectral simulation program that includes j-specific electric dipole transitions, a Greens Function modelling of the electronic structure and transitions into both L+1 and L-1 channels, where L is the angular momentum of the initial state electron. The Density of States (DOS) is an orbital angular momentum (L) projected density of states, which does not include total angular momentum specificity (J). Hence there is no proper J dependence and the spin-orbit splitting is neglected. Angular momentum coupling is neglected.

FEFF simulations (Table 1) of the Ce $M_{4,5}$ XAS transitions and the corresponding DOS (Figure 5) were carried out using a very simple $CeO_8$ cluster that mimicked the local geometry of the Ce in $CeO_2$. It has been found empirically that freezing the f states and allowing f state mixing with the other manifolds can reproduce the band gap of an oxide qualitatively, as illustrated in Figure 5. [52] In this case, the Fermi Edge is pinned to the Valence Band maximum, but it is noteworthy that these simplistic FEFF calculations reproduce the results of far more sophisticated calculations, [53-56] at least qualitatively and perhaps even semi-quantitatively.

Table 1 The $CeO_8$ $M_{4,5}$ XAS FEFF parameters were as follows:

| | | |
|---|---|---|
| (1)XANES 9, 0.07, 0.0; FMS 3.0; | (4)Exchange off; | (7)CHWIDTH 3; |
| (2)SCF 30,0,100,0.05,10; | (5)Core Hole on, No; | (8)Absolute Cross Section; |
| (3)UNFREEZE; OFF | (6)SO2 off; | (9)LDOS -40, 10, 0.05 |

Notes

(1) The XANES routine is used when a calculation of the near edge structure including the atomic background and absolute energies is desired (max, step size far from edge, step size near the edge). FMS: Compute the full multiple scattering within a sphere of the specified radius.
(2) Controls FEFF's automated self-consistent potential calculation. (30 = cluster radius, 0 specifies a solid, 100 = maximum number of iterations, 0.2 = convergence accelerator factor, 1 = maximum number of iterations with the mixing algorithm)
(3) Unfreeze-OFF = f electrons are frozen and do not participate in the SCF convergence
(4) The energy dependent exchange correlation potential is off.
(5) There are three options for the treatment of the core hole: the simple non-screened Final State Rule (FSR); a screened core hole (RPA) and calculate without a core hole (No).
(6) The amplitude reduction factor-OFF
(7) Sets the core-hole lifetime manually, instead of using the internal table.
(8) Disables the default normalization and keeps the absolute values.
(9) L specific DOS, from -40 eV to 10 eV, with 0.05 eV broadening

*Figure 5 (Below) Presented here are the L-specific DOS results for the $CeO_8$ Cluster FEFF. The Ce-DOS are vertically positive deflections and the Oxygen -DOS are vertically negative deflections. The Occupied Density of States (ODOS) are at negative energies and the Unoccupied Density of States (UDOS) at positive energies. $E_F$ is the Fermi Energy, at zero eV. The DOS results were convolved with a Lorentzian broadening function with a Half-Width-at-Half-Height (HWHH) of 0.05 eV. There is small Ce 4f ODOS peak just below the Fermi Energy.*

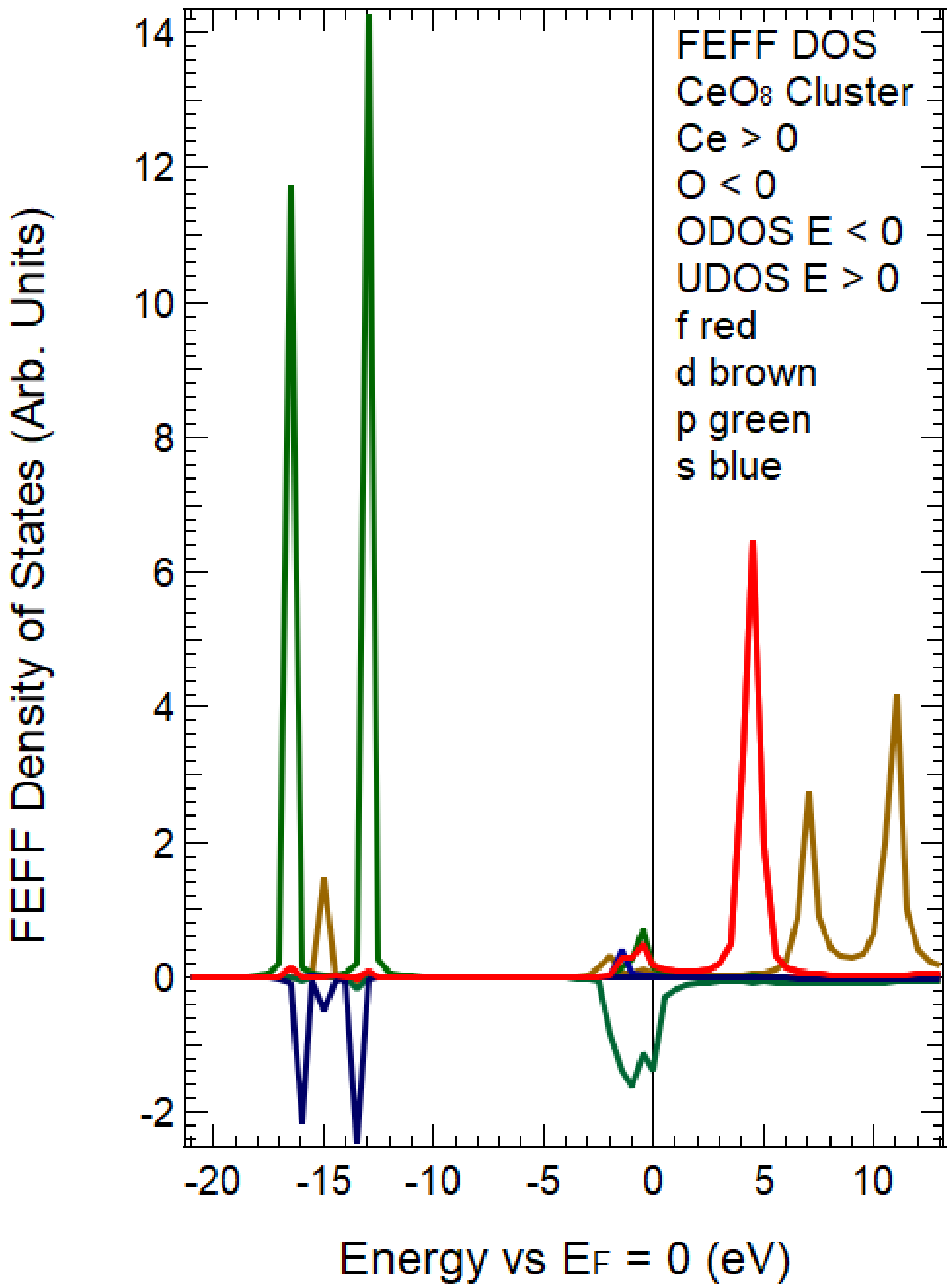

Figure 5

## III Results and Discussion

There are two new pieces of evidence that support the hypothesis that the Ce $M_5$ RIPES is driven by a Ce 5d connection.

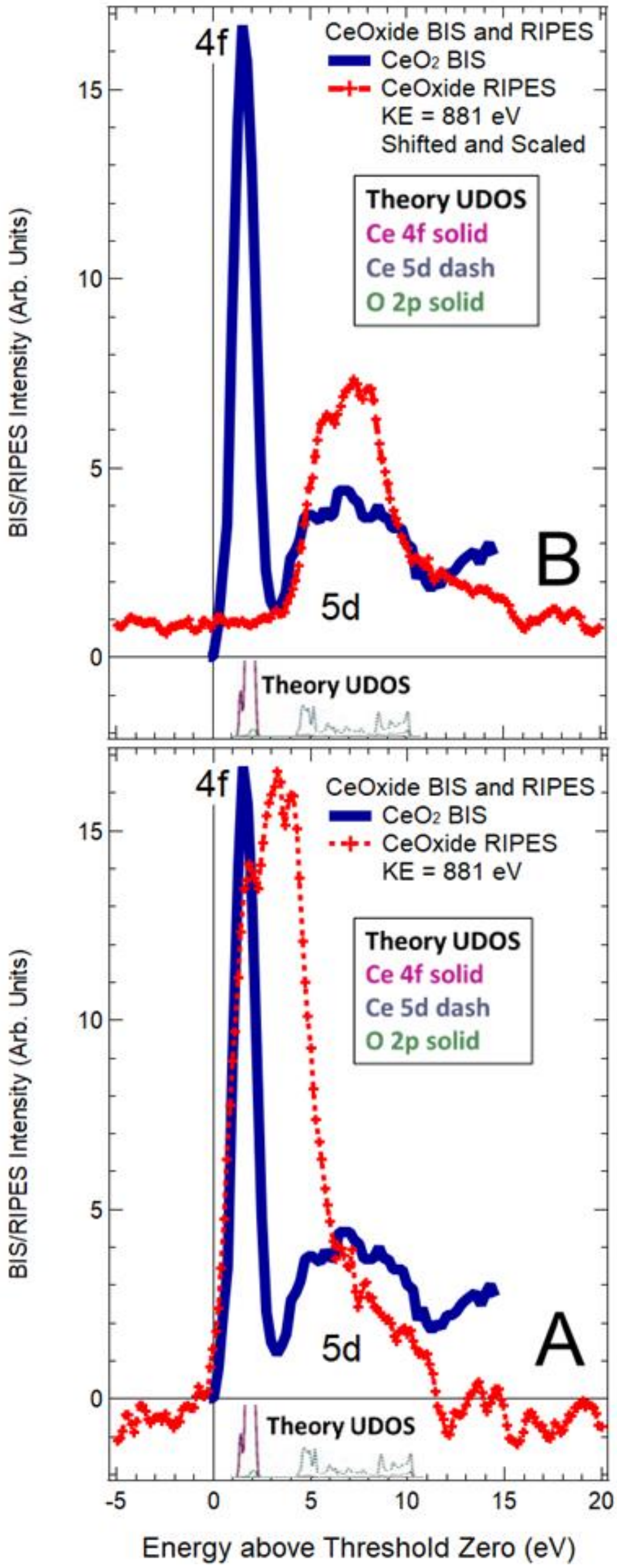

*Figure 6 Shown here is the comparison of the BIS of $CeO_2$ from Ref [19, 22] (blue solid line) and the RIPES of cerium oxide (red plus and dashed line). A: alignment on the threshold. B: alignment on the Ce 5d peak. The theoretical predictions of the placement of the Ce 4f, Ce5d and O 2p states is also shown in A and B [54], confirming the peak assignments. Note the agreement in Panel B between the 5d fine structure in the BIS and the fine structure in the RIPES.*

# Resonant Inverse Photoelectron Spectroscopy of Ce and U Oxides

## IIIa Comparison to BIS

There have been at least two reports of investigations of $CeO_2$ using Bremsstrahlung Isochromat Spectroscopy (BIS). [19, 22, 23] They both produced very similar BIS spectra and each is significantly different than the RIPES of cerium oxide, as illustrated in Panel A of Figure 6. However, shifting the RIPES to align the RIPES with the second smaller peak produces a remarkable agreement, down to the fine structure inside the peak. This agreement strongly supports the hypothesis that the Ce $M_5$ RIPES process is going through the Ce 5d states.

## IIIb CE $M_{4,5}$ XAS

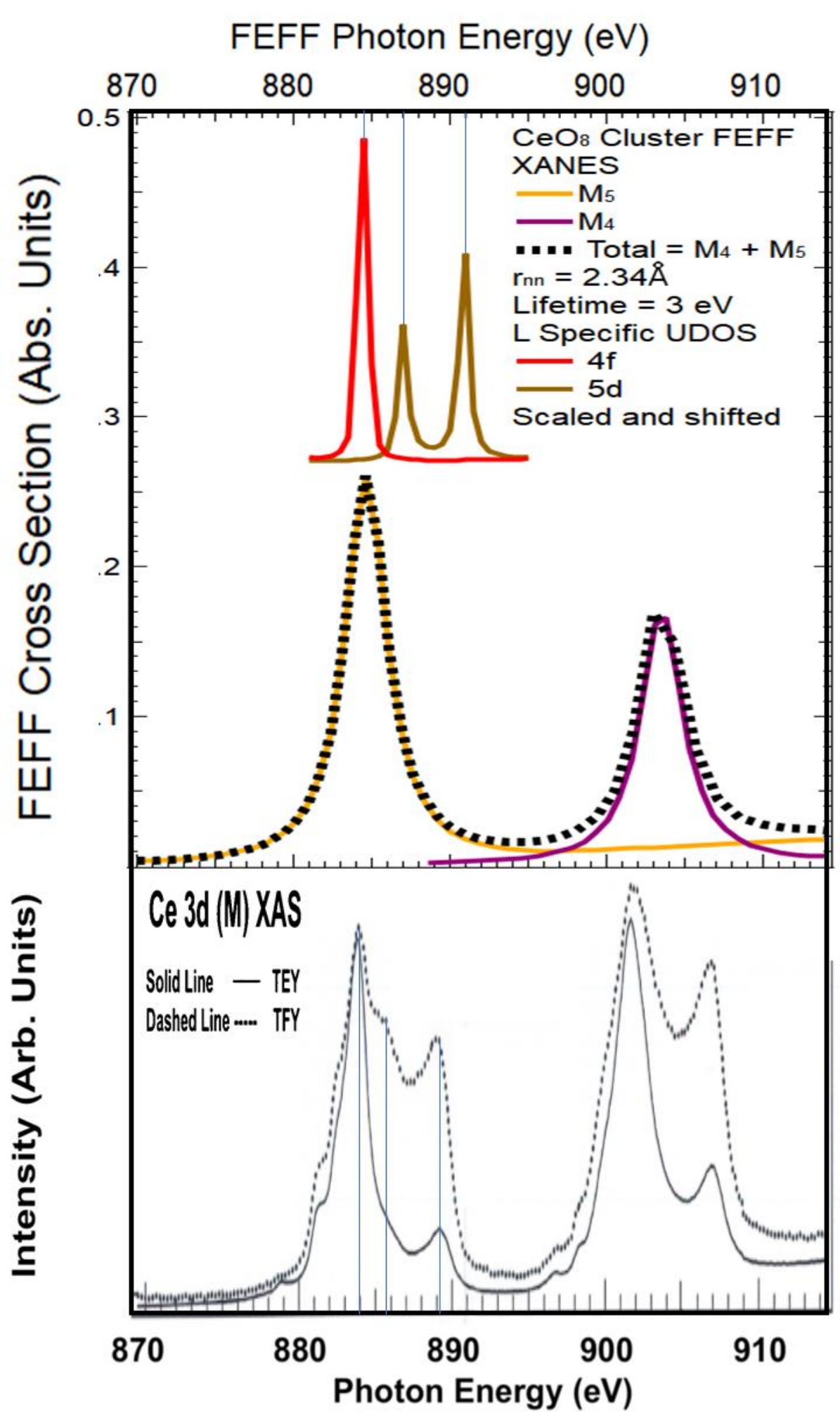

Figure 7
Presented here is the comparison of the experimental spectra and FEFF simulations for the Ce $M_{4,5}$ (3d) XAS transitions in cerium oxide. The FEFF model is a $CeO_8$ cluster, consistent with the local environment around a Ce in a $CeO_2$ sample. TEY is total electron yield. TFY is total fluorescence yield. TEY is more surface sensitive. TFY is more bulk sensitive. Data is from References [27,28].

Another avenue to address the question of Ce 5d is to separately probe the individual steps of the processes illustrated in Panel A of Figure 2. Specifically, one should look at the steps in the case of the Indirect Channel of the Satellite peak. Consider the XAS spectra in Figure 7. [27, 28] These transitions would correspond to the excitation of a core level electron into the conduction band. While the sample used in this study is defined as cerium oxide, the Total Electron Yield (TEY) spectrum in Figure 7 is in excellent agreement with spectra reported by Kaindl et al for $CeO_2$, also using total electron yield. [57] The single particle, electric dipole prediction of FEFF as well as the underlying UDOS are also shown. Clearly, the major peak in each FEFF XAS spectrum is the simple transition of a single electron from the 3d core level into the empty 4f states. The small fine structure on the main peaks is due to total angular momentum coupling. [27, 28, 58] However, the larger additional features may correspond to Ce 5d UDOS. While the RIPES should exhibit a shift between the main and satellite peaks of 2Δ (Panels A and B, Figure 2), the separate excitation step should only have a separation of a single Δ, where Δ is the separation of the f and d manifolds. (Panel A, Figure 2) [26] This assignment has two major problems. (1) it is inconsistent with the earlier assignment of the main peak as $3d^9 4f^2$ and the satellite as $3d^9 4f^1$. [57, 59, 60] (2) Preliminary testing with FEFF indicates that the higher order transitions, electric quadrupole and magnetic dipole, have negligible intensities. However, it is still of use to note the correlation.

## IV Conclusion and Summary

The soft x-ray Resonant Inverse Photoelectron Spectroscopy (RIPES) of the $M_5$ transition in cerium oxide and the $N_5$ transition in uranium dioxide and supporting instrumentation has been discussed. The best evidence to support the hypothesis that the Ce 5d states are intrinsic to the resonance process in cerium oxide is the very strong agreement between the Ce5d structure in BIS with the RIPES peak. The strong correlation between the Ce $M_{4,5}$ (3d) XAS and the $CeO_8$ Ce 5d UDOS is intriguing but contrary to conventional interpretation and not supported by modelling, at least as of now.

ACKNOWLEDGEMENTS: The author wishes to thank the University of Wisconsin-Oshkosh and its faculty and staff for their ongoing support and encouragement. It is a gem set beside the Fox River and Lake Winnebago.

DATA AVAILABILITY: The data that support the findings of this study are available from the corresponding author upon reasonable request.

AUTHORS DECLARATION: The author has no conflicts to disclose.